\documentclass[%
 aip,
 amsmath,amssymb,
 reprint,%
]{revtex4-1}

\usepackage{graphicx}
\usepackage{dcolumn}
\usepackage{bm}

\usepackage[utf8]{inputenc}
\usepackage[T1]{fontenc}
\usepackage{mathptmx}
\usepackage{stmaryrd}
\usepackage{color}

\begin{document}

\preprint{AIP/123-QED}

\title{Leidenfrost drop impact on inclined superheated substrates}

\author{Yujie Wang} 
\affiliation{Center for Combustion Energy, and Key Laboratory for Thermal Science and Power Engineering of Ministry of Education, and Department of Energy and Power Engineering, Tsinghua University, 100084 Beijing, China}
\author{Ayoub El Bouhali} 
\affiliation{Center for Combustion Energy, and Key Laboratory for Thermal Science and Power Engineering of Ministry of Education, and Department of Energy and Power Engineering, Tsinghua University, 100084 Beijing, China}
\author{Sijia Lyu}
\email{lvsj16@mails.tsinghua.edu.cn}
\affiliation{Center for Combustion Energy, and Key Laboratory for Thermal Science and Power Engineering of Ministry of Education, and Department of Energy and Power Engineering, Tsinghua University, 100084 Beijing, China}
\author{Lu Yu}
\affiliation{State Key Laboratory of Hydro Science and Engineering, and Department of Energy and Power Engineering, Tsinghua University, 100084 Beijing, China}
\author{Yue Hao}
\affiliation{Center for Combustion Energy, and Key Laboratory for Thermal Science and Power Engineering of Ministry of Education, and Department of Energy and Power Engineering, Tsinghua University, 100084 Beijing, China}
\author{Zhigang Zuo}
\affiliation{State Key Laboratory of Hydro Science and Engineering, and Department of Energy and Power Engineering, Tsinghua University, 100084 Beijing, China}
\author{Shuhong Liu}
\affiliation{State Key Laboratory of Hydro Science and Engineering, and Department of Energy and Power Engineering, Tsinghua University, 100084 Beijing, China}
\author{Chao Sun} 
\affiliation{Center for Combustion Energy, and Key Laboratory for Thermal Science and Power Engineering of Ministry of Education, and Department of Energy and Power Engineering, Tsinghua University, 100084 Beijing, China}
\affiliation{Department of Engineering Mechanics, and School of Aerospace Engineering, Tsinghua University, Beijing 100084, China}

%

\date{\today}

\begin{abstract}
In real applications, drops always impact on solid walls with various inclinations. For the oblique impact of a Leidenfrost drop, which has a vapor layer under its bottom surface to prevent its direct contact with the superheated substrate, the drop can nearly frictionlessly slide along the substrate accompanied by the spreading and the retracting. To individually study these processes, we experimentally observe ethanol drops impact on superheated inclined substrates using high-speed imaging from two different views synchronously. We first study the dynamic Leidenfrost temperature, which mainly depends on the normal Weber number ${We}_\perp $. Then the substrate temperature is set to be high enough to study the Leidenfrost drop behaviors. During the spreading process, drops always keep uniform. And the maximum spreading factor $D_m/D_0$ follows a power-law dependence on the large normal Weber number ${We}_\perp $ as $D_m/D_0 = \sqrt{We_\perp /12+2}$ for {\it We}$_\perp \geq 30$. During the retracting process, drops with low impact velocities become non-uniform due to the gravity effect. For the sliding process, the residence time of all studied drops is nearly a constant, which is not affected by the inclination and {\it We} number. The frictionless vapor layer results in the dimensionless sliding distance $L/D_0$ follows a power-law dependence on the parallel Weber number {\it We}$_\sslash$ as $L/D_0 \propto We_\sslash^{1/2}$. Without direct contact with the substrate, the behaviors of drops can be separately determined by ${We}_\perp $ and {\it We}$_\sslash$. When the impact velocity is too high, the drop fragments into many tiny droplets, which is called the splashing phenomenon. The critical splashing criterion is found to be {\it We}$_\perp ^*\simeq$ 120 or $K_\perp  = We_\perp Re_\perp^{1/2} \simeq$ 5300 in the current parameter regime.

\end{abstract}

\maketitle

\section{INTRODUCTION}

Drop impact dynamics on a substrate has a wide range of industrial applications, such as spray combustion~\cite{Moreira2010Advances}, spray cooling~\cite{Kim2007Spray}, inkjet printing~\cite{derby2010inkjet}, spray coating~\cite{andrade2013drop} and so on. The perpendicular drop impact on solid substrates has been widely studied to reveal the complicated dynamic mechanisms between the drop and substrate~\cite{josserand2016drop,rioboo2001outcomes,gunjal2005dynamics,tran2012drop}. However, in practical applications, the drop usually impacts on the substrate obliquely. In these cases, besides spreading and rebounding, the drop slides along the inclined substrate. And the attachment of drop to the substrate causes different impact regimes~\cite{yeong2014drop,zhang2017drop,vsikalo2005impact}, such as deposition, rivulet, sliding, rolling, partial rebound, and complete rebound~\cite{antonini2014oblique}. Even for superhydrophobic substrates, the effect of the substrate is still very important\cite{zhang2017drop}. Then we want to know if there is no attachment between the drop and the substrate, how the drop behaves when it impacts on the inclined substrate. 

A very ideal drop is the Leidenfrost drop, which has a vapor layer under its bottom surface to prevent its direct contact with the heated substrate~\cite{leidenfrost1756aquae,quere2013leidenfrost}. Thanks to the occurrence of the vapor layer, the drop can move freely along the substrate~\cite{vakarelski2011drag, saranadhi2016sustained,yu2007multi}. Through this way, we can more clearly understand the effects of the impact velocity and the inclination on the drop impact dynamics. These results can help us to adjust the range of various parameters in industrial applications. In addition, the oblique drop impact is an easy way to alter the moving direction and the shape of drops, which can be applied in the spray fields.

\section{EXPERIMENTAL METHODS} \label{exp}

\begin{figure}
\centering
	\includegraphics[width=1 \linewidth]{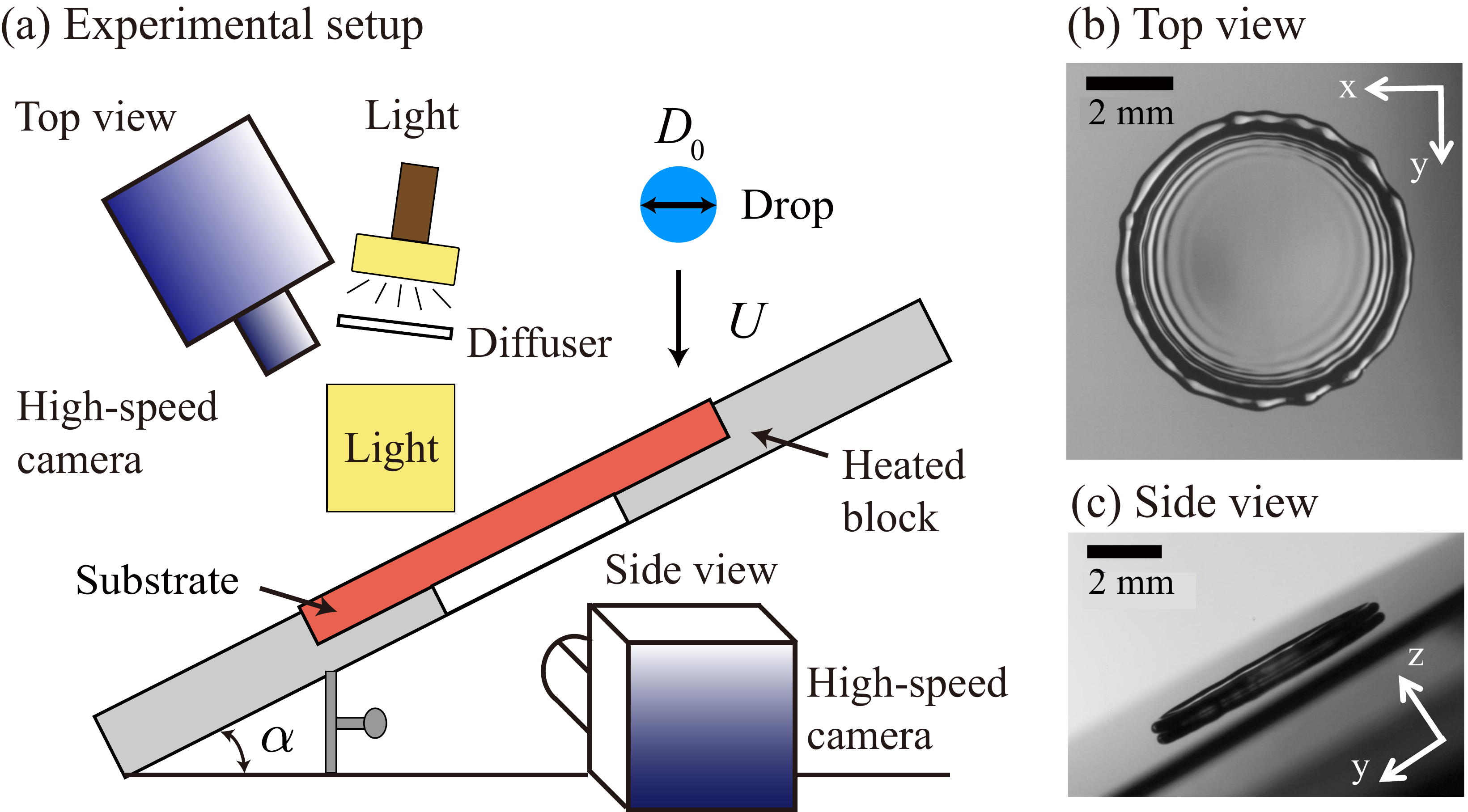}  
\caption{Sketch of the experimental setup. An ethanol drop with an initial diameter $D_0$ $\approx$ 2.1 mm impacts on a heated substrate with an inclination $\alpha$. A polished silicon wafer is used as the target substrate. In order to investigate the dynamic Leidenrost temperature $T_L$, we use a transparent sapphire substrate to record the bottom view observation. The substrate is placed on a heated aluminum block, which is heated by four heating rods and the temperature is controlled by a PID controller. Two sets of high-speed cameras are placed in two different directions. The corresponding images recorded from the top and the side views are shown in Fig.~\ref{setup} (b, c).}
\label{setup}
\end{figure}

The sketch of the experimental setup is shown in Fig.~\ref{setup}(a). An ethanol drop is generated from the tip of a blunt needle connected to a syringe pump (Harvard Apparatus PHD ULTRA) with a constant flow rate ($ \approx $ 0.08 ml/min). The initial diameter of the drop $D_0$ is 2.1~$\pm$~0.1~mm. We change the height of the needle to adjust the impact velocity of the drop $U$, within the range of 0.2 to 3.0 m/s. The corresponding Weber number {\it We}$ = \rho D_0 U^2/\sigma$ ranges from 3 to 680, where $\rho$ is the density of ethanol and $\sigma$ the surface tension of ethanol at room temperature respectively. Two sets of high-speed cameras (Photron FASTCAM Mini AX200 \& Photron FASTCAM Mini UX100) with macro lenses (Canon EF 24-105 mm) are used to record the drop behaviors from top and side views synchronously with a frame rate of 6,400 fps. Two halogen lamps are used to supply the reflected light for the top view and the backlight for the side view. Fig.~\ref{setup}(b) shows a snapshot recorded from the top view, which displays the spreading and retracting processes of the drop. Fig.~\ref{setup}(c) shows a snapshot recorded from the side view, which records the sliding process of the drop. 

In our experiments, a smooth silicon wafer is used as the target substrate (the average surface roughness $\approx$ 10 nm). The diffuser scatters the light on the non-transparent substrate to facilitate clear observations from the top view. The substrate is placed on a heated aluminum block, which is heated by four heating rods. The temperature of the substrate $T_s$ is controlled by a PID controller with an accuracy of $\pm 0.5 ^\circ$C. In addition, in order to get the accurate Leidenfrost temperature, we use a transparent sapphire as another substrate. A 20 mm $\times$ 50 mm rectangular hole is left at the block center, forming an observation window for the bottom view. A long-working-distance microscope with a coaxial LED lamp is placed in the bottom view. Combining side and bottom views, we can identify the boiling characteristics of drops and obtain the accurate Leidenfrost temperature $T_L$. Then we choose a high enough substrate temperature to make sure all studied drops stay in Leidenfrost state.

\section{RESULTS AND INTERPRETATION} \label{results}

We divide results and interpretation into five parts, including the dynamic Leidenfrost temperature, spreading dynamics, residence time, sliding distance, and splashing criterion. 

\subsection{Dynamic Leidenfrost temperature} 

\begin{figure}
\centering
\includegraphics[width=1 \linewidth]{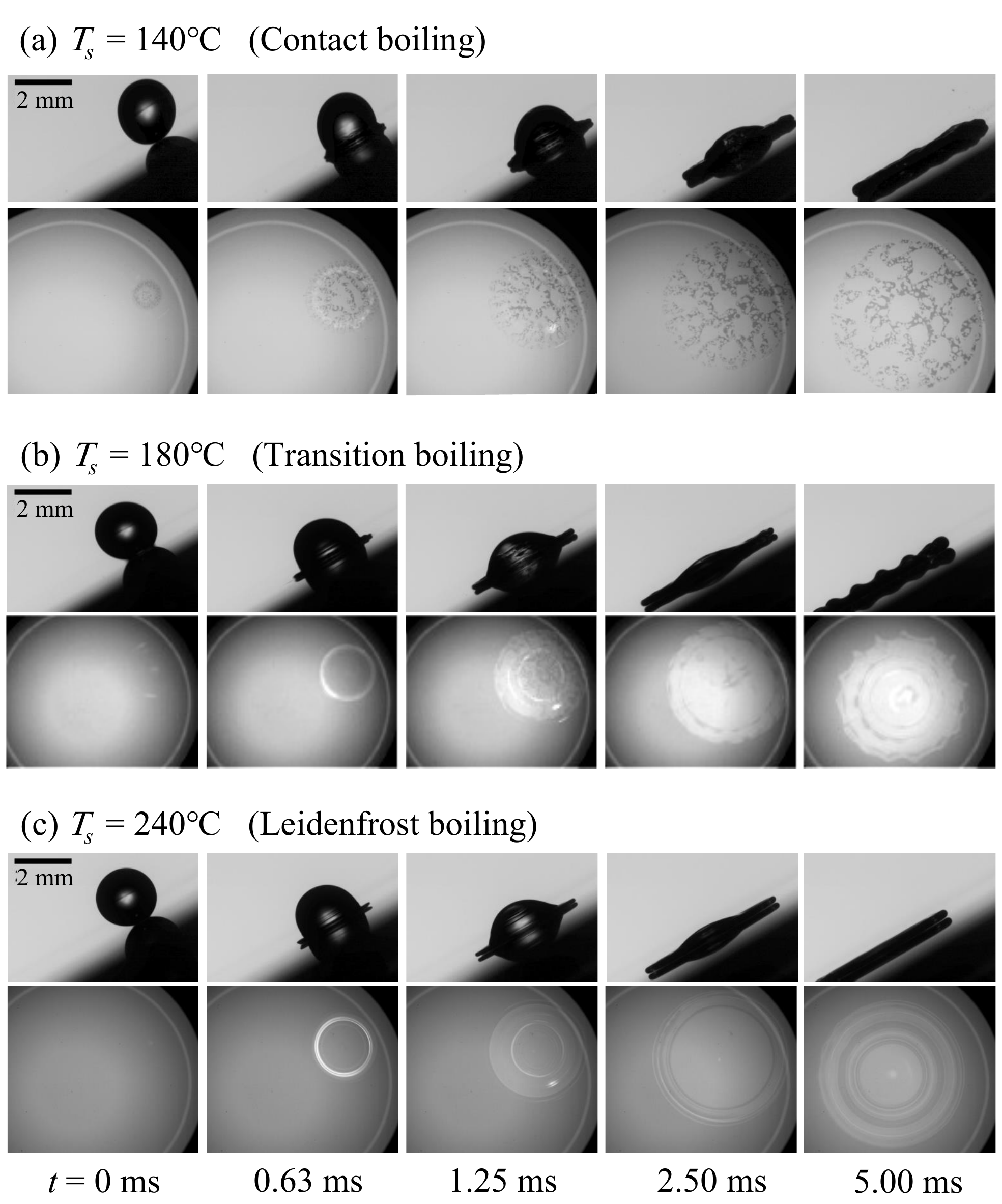}
\caption{Snapshots of ethanol drops impacting on an inclined sapphire substrate ($\alpha$ = 30$^\circ$) at different substrate temperatures. Drops stay in three different boiling regimes, such as contact boiling, transition boiling, and Leidenfrost boiling. The impact velocity $U$ is 1 m/s and Weber number $We$ is 75 for these three cases. In each group, the time sequence is 0, 0.63, 1.25, 2.50, and 5.00 ms. The images in the upper row are recorded from the side view, and lower ones are recorded from the bottom view. The dark spots in the bottom-view images indicate the liquid wetting area. The scale bars represent 2 mm.}
\label{regime}
\end{figure}

When a drop perpendicularly impacts on heated solid substrates, there are three different boiling regimes, i.e., contact boiling, transition boiling, and film boiling (Leidenfrost state)~\cite{Tran2012prl,Shirota2016}. These three different boiling regimes are also observed for the inclined drop impact. In order to accurately determine the Leidenfrost temperature $T_L$ (the minimum temperature of the film boiling regime), we use a transparent sapphire base as a target substrate. Combining the side and the bottom views, we can determine the accurate regime of a drop. Figure~\ref{regime} shows the side-view (upper) and bottom-view (lower) recordings for a drop impact on a plate with an inclination of 30 degrees. In the contact boiling regime (see Fig.~\ref{regime}(a)), the drop directly contacts with the substrate, as shown in the dark area from the bottom view. As the substrate temperature increases, more vapor is generated under the drop bottom, therefore the drop only partially wets the substrates (see the dark area in the third image from the bottom view). And the drop stays in an unsteady state as shown in the side view. This regime is called the transition boiling (see Fig.~\ref{regime}(b)). When the substrate temperature increases beyond a critical temperature, enough vapor is generated to elevate the drop from the substrate during the whole impact process; this is called the Leidenfrost regime (see Fig.~\ref{regime}(c)). The transition temperature between the transition boiling regime and the Leidenfrost regime is defined as the dynamic Leidenfrost temperature ($T_L$).

\begin{figure}
\centering
	\includegraphics[width=0.9 \linewidth]{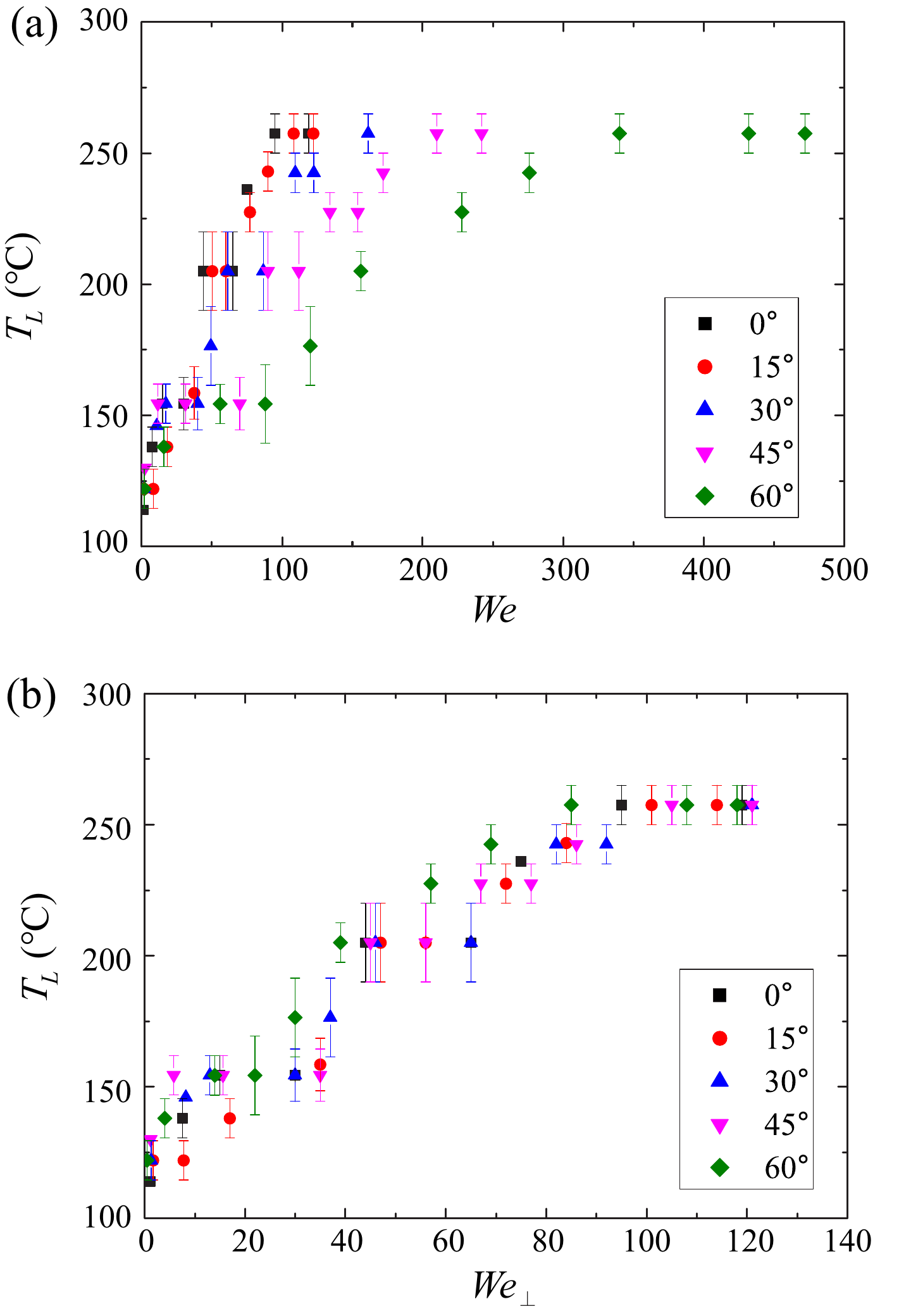}
 \caption{Dynamic Leidenfrost temperature $T_L$ (the minimum temperature to maintain the drop in the Leidenfrost regime) versus (a) Weber number {\it We} (b) normal Weber number {\it We}$_\perp $ for different inclinations.}
       \label{TL}
\end{figure}

Then we change the inclination $\alpha$ of the substrate and want to find the dependence of the dynamic Leidenfrost temperature on {\it We} number for various $\alpha$. In Fig.~\ref{TL}(a), for a fixed inclination $\alpha$, the corresponding dynamic Leidenfrost temperature $T_L$ increases with {\it We}. When {\it We} is high enough, $T_L$ becomes insensitive with $We$. This dependence can be rationalized by comparing the inertial pressure of the drop and the vapor pressure. The inertial pressure of the drop is proportional to {\it We} and the vapor pressure increases with $T_s$. Thus, in order to maintain a drop at a high {\it We} in the Leidenfrost regime, a high substrate temperature is required to produce enough vapor to overcome a large dynamic pressure of the impacting drop. The inclination angle $\alpha$ in the study is $0^\circ$, $15^\circ$, $30^\circ$, $45^\circ$, and $60^\circ$. For a fixed {\it We}, increasing the inclination can effectively decrease  $T_L$.  For example, compared to the horizontal substrate case, $T_L$ for the inclined substrate with $\alpha = 60^\circ$ reduces roughly 100$^\circ$C when {\it We} is about 100. The physical reason for the reduction of dynamic Leidenfrost temperature on the inclined substrates is due to the reduction of the `effective impacting velocity', which is the normal velocity to the substrate $U_\perp  = U \cos\alpha$. Using the perpendicular velocity, we define the normal Weber number as

\begin{equation}
We_\perp  = \frac{\rho D_0 U_\perp ^2}{\sigma}.
\end{equation}

As shown in Fig.~\ref{TL}(b), all $T_L$ evolutions of the {\it We}$_\perp $ for different inclinations collapse to a master curve, which means the dynamic Leidenfrost temperature mainly depends on the normal impact velocity. This finding indicates that it's possible to reduce the dynamic Leidenfrost temperature of an impacting drop by simply inclining substrates due to the reduction of the normal impact velocity.

In addition, from Fig.~\ref{TL}(a), we know that 300$^\circ$C is higher than all Leidenfrost temperatures for our studied {\it We} range. Because the silicon wafer has a similar roughness as that of the sapphire wafer, we set $T_s=$300$^\circ$C to study Leidenfrost drops' behaviors on the silicon wafer below.

\subsection{Leidenfrost regime: spreading dynamics} \label{exp2}

\begin{figure*}
\centering
\includegraphics[width=0.9 \linewidth]{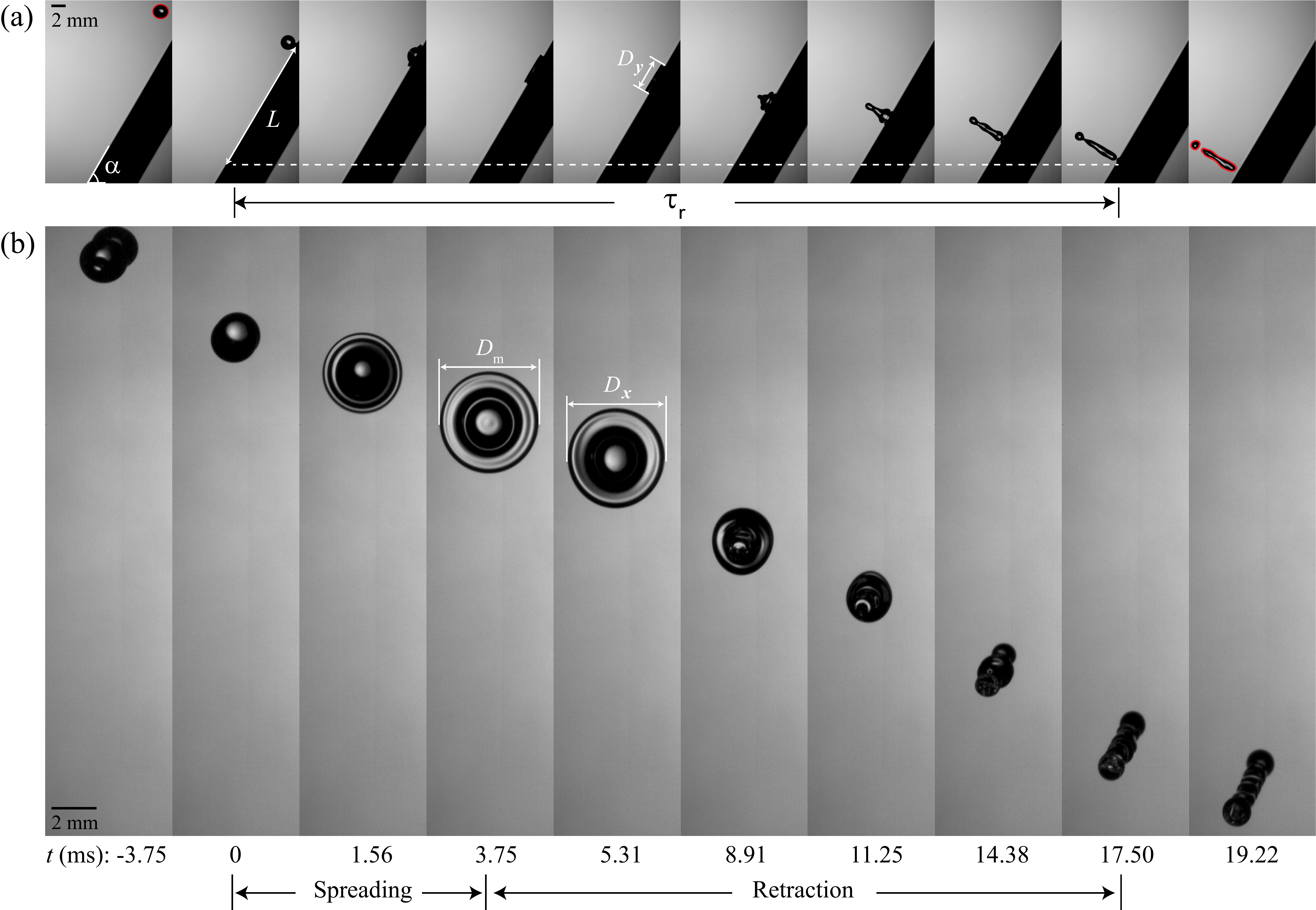}
\caption{An entire recording of a Leidenfrost drop impacting on an inclined substrate. The setting parameters are $\alpha = 60^ \circ$, $T_s = 300^ \circ$C, and {\it We}=15. (a) The side-view recording shows the definitions of the inclination $\alpha$, the sliding distance $L$, the diameter $D_y$ in the $y$ direction, and the residence time $\tau_r$. (b) The top-view recording shows the definitions of the maximum spreading diameter $D_m$, the diameter $D_x$ in the $x$ direction, and the spreading and the retracting processes. When the drop contacts the substrate, this moment is defined as $t$ = 0. When the drop bounces off the substrate, this moment is defined as the end of the residence time. After bouncing from the substrate, the sphere drop becomes nearly cylindrical (see red lines). Scale bars represent 2 mm.}
\label{entire}
\end{figure*}

\begin{figure}
\centering
\includegraphics[width=0.9 \linewidth]{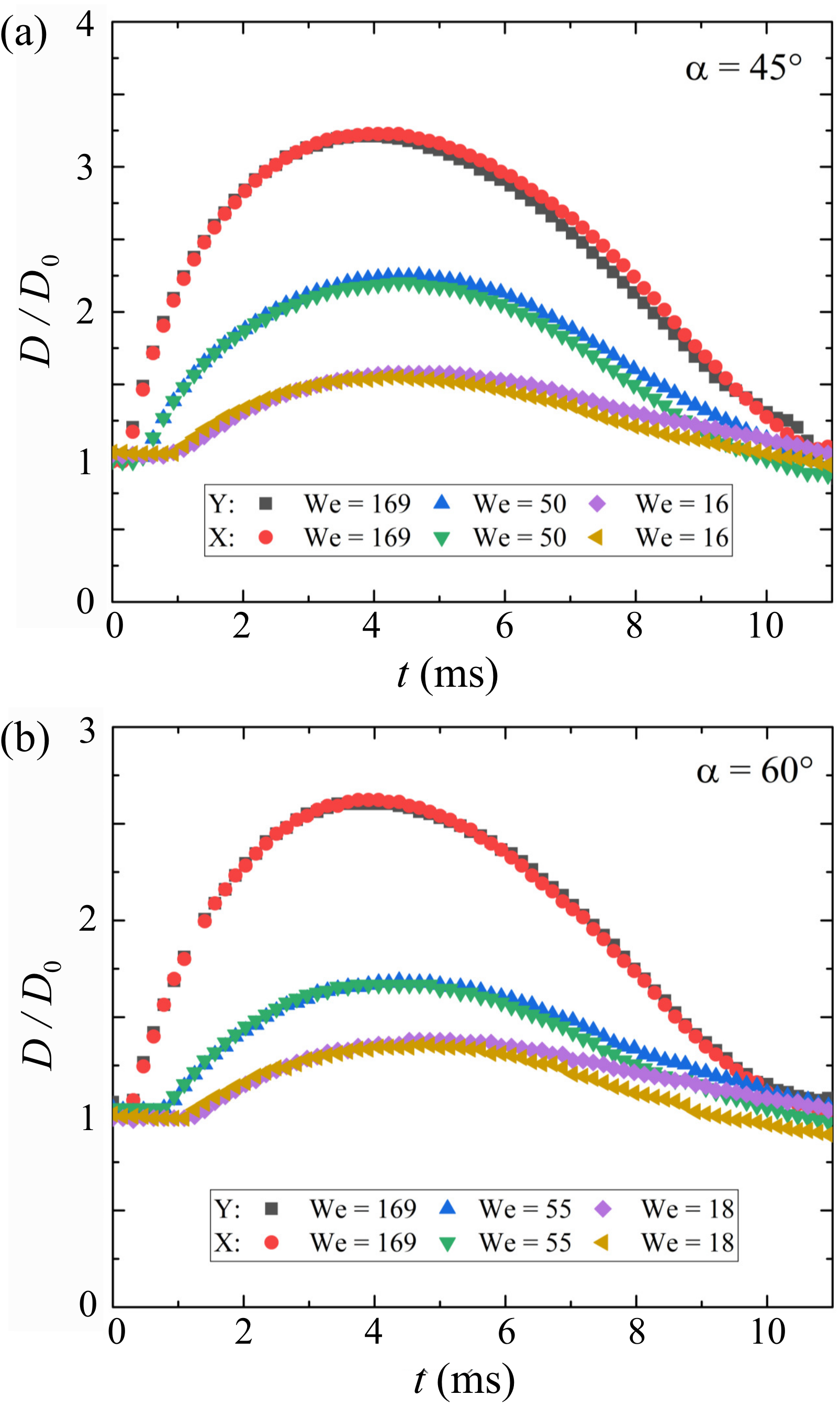}
\caption{ The dimensionless spreading factor $D/D_0$ in the $x$ and $y$ directions versus time $t$  for two different inclinations. (a) $\alpha$ = 45$^{\circ}$ and (b) $\alpha$ = 60$^{\circ}$. The initial drop diameter $D_0$ is 2.1 mm. 
}
\label{D}
\end{figure}

When a drop impacts on a horizontal substrate, the drop is nearly axisymmetric. However, when the drop impacts upon an inclined substrate at room temperature, it is found that the drop has a non-axisymmetric spread and retraction due to the attachment to the substrate~\cite{Choi2016Drop}. For a Leidenfrost drop, it is not directly affected by the substrate. But due to the effect of gravity, the spreading and retracting dynamics of the drop may be changed.

Figure~\ref{entire} shows the entire process of a Leidenfrost drop impacting on an inclined substrate. The setting parameters are $\alpha = 60^ \circ$, $T_s = 300^ \circ$C, and {\it We} = 15. In Fig.~\ref{entire}(a), the side-view recording shows the definitions of the inclination $\alpha$, the sliding distance $L$, the diameter $D_y$ in the $y$ direction, and the residence time $\tau_r$. In Fig.~\ref{entire}(b), the top-view recording shows the definitions of the maximum spreading diameter $D_m$, the diameter $D_x$ in the $x$ direction, and the spreading and the retracting processes. After bouncing from the substrate, the sphere drop becomes nearly cylindrical, which has a larger surface area (see red lines in the first and the last snapshots of Fig.~\ref{entire}(a)).

Figure~\ref{D} shows a quantitative study of $D_x/D_0$ and $D_y/D_0$ versus time for two different inclinations $45^\circ$ and $60^\circ$ at three different {\it We} respectively. For all cases, drops always stay in the Leidenfrost regime. The impacting process can be recognized as two stages, i.e. the spreading process and the retracting process. As shown in the Fig.~\ref{D}, the temporal evolutions of the diameter in $x$ and $y$ directions during the spreading process are very close, which means the inclination of the substrate and the $We$ number do not induce an azimuthal variation of the Leidenfrost drop during the spreading process. 
In addition, the maximum diameters are nearly the same in two directions. As expected, it increases with {\it We} for a fixed inclination. 
Comparing Fig.~\ref{D}(a) with Fig.~\ref{D}(b), we find the maximum spreading diameter increases with the decreasing inclination at a fixed {\it We} $=169$. It suggests that a Leidenfrost drop can spread uniformly in two directions on inclined substrates, and the maximum spreading diameter is strongly affected by the impact velocity and the substrate inclination. However, during the retraction process, it shows a slight difference in these two directions, especially for a lower {\it We} or a higher inclination. The drop retracts a little more slowly in the $y$ direction due to the gravity effect.

\begin{figure}
\centering
\includegraphics[width=0.9 \linewidth]{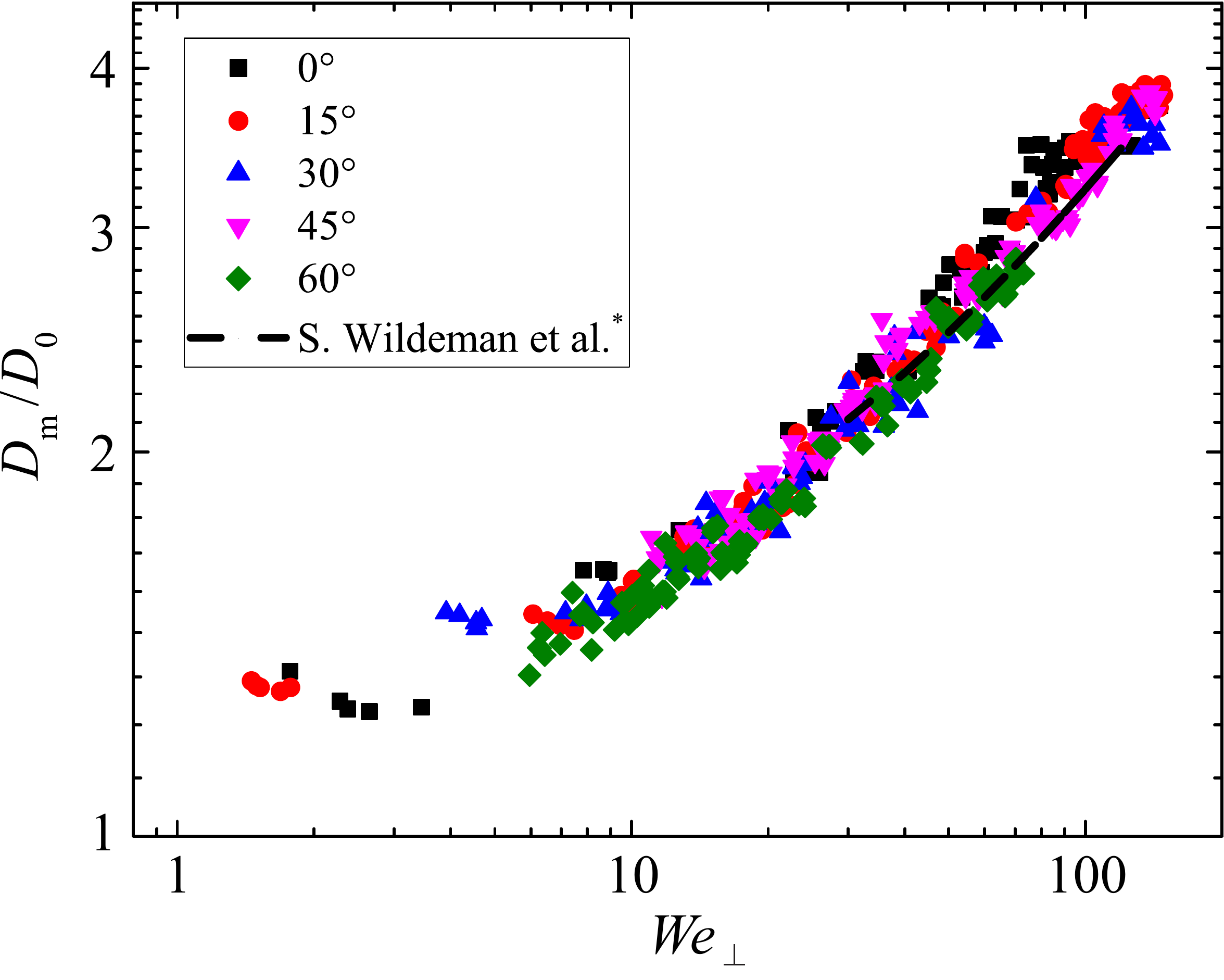}
\caption{The maximum spreading factor $D_m/D_0$ verus the normal {\it We}$_\perp $ for different inclinations. $D_m/D_0$ is mainly determined by {\it We}$_\perp $. Our experimental results also fit well with the theoretical model~\cite{wildeman2016spreading} at high {\it We}$_\perp $ ({\it We}$_\perp  \geq$ 30), which is shown as the black dashed line.}
\label{Dm}
\end{figure}

We now investigate the quantitative dependence of the maximum spreading diameter $D_m$ on {\it We} number. In Fig.~\ref{Dm}, we show a log-log plot of the maximum spreading factor $D_m / D_0$ versus {\it We}$_\perp $ for various inclinations. All spreading factors coalesce to a master curve. It implies the maximum spreading factor mainly depends on the normal Weber number {\it We}$_\perp $. $D_m / D_0$ increases with {\it We}$_\perp $, which means a larger effective normal impact velocity induces a larger spreading diameter. When we incline the substrate, the effective normal impact velocity decreases, which induces the reduction of $D_m / D_0$, suggesting less initial kinetic energy of the drop is transformed to its surface energy. Thus, we can decrease the maximum spreading diameter by increasing the inclination of the substrate. In addition, we compare our experimental results with the theoretical model for drop impact on the free-slip surface proposed by S. Wildeman et al.~\cite{wildeman2016spreading}. The theoretical model of the dimensionless maximum spreading diameter $D_m/D_0 = \sqrt{We_\perp /12+2}$ for {\it We}$_\perp \geq 30$, which is shown as the black dashed line in Fig.~\ref{Dm}. Our experimental results agree well with the theoretical model at high {\it We}$_\perp $ numbers ({\it We}$_\perp \geq 30$), which means the vapor layer can be treated as the free-slip surface. Thus, we conclude that, for the Leidenfrost drop impact upon an inclined substrate, the impact velocity can be decomposed to a normal component and a parallel component to the substrate. And the spreading dynamics is strongly dependent on the normal velocity.

\subsection{Leidenfrost regime: residence time} \label{exp2}
 
When a Leidenfrost drop impacts upon a superheated substrate, it first deforms and radially spreads on the substrate. We define the moment that drop contacts the substrate is the start moment of the contact time. When the impact velocity is not too high, the integral drop bounces off the substrate. Otherwise, it fragments into a lot of tiny drops (splashing). Here, we focus on the regime where the drop spreads and bounces off the substrate without the splashing. The moment that drop bounces off the substrate is the end moment of the contact time. The period between the start and the end of the contact time is the residence time $\tau_r$. For drop impact on substrates at room temperature, the attachment between the drop and the substrate causes the drop becomes non-axisymmetric and then reduces the residence time of the drop~\cite{zhang2017drop}. So we want to know if there is no attachment between the drop and the substrate, what parameters determine the residence time of a Leidenfrost drop. 
Normally, the residence time of a drop impact on an unheated superhydrophobic substrate or a horizontal superheated substrate is connected to that of a freely oscillating drop~\cite{Richard2002,Tran2013softmatter}. In the limit of low viscosity, the residence time can be obtained by balancing the inertia force with the capillary force.
With the prefactor calculated by Rayleigh~\cite{rayleigh1879capillary}, the residence time of a freely oscillating drop is given as: 
\begin{equation}
 \tau_0 = \dfrac{\pi}{4}\sqrt{\dfrac{\rho_b\emph{D}_0^3}{\sigma_b}}.
\end{equation}

\begin{figure}
\centering
\includegraphics[width=0.95 \linewidth]{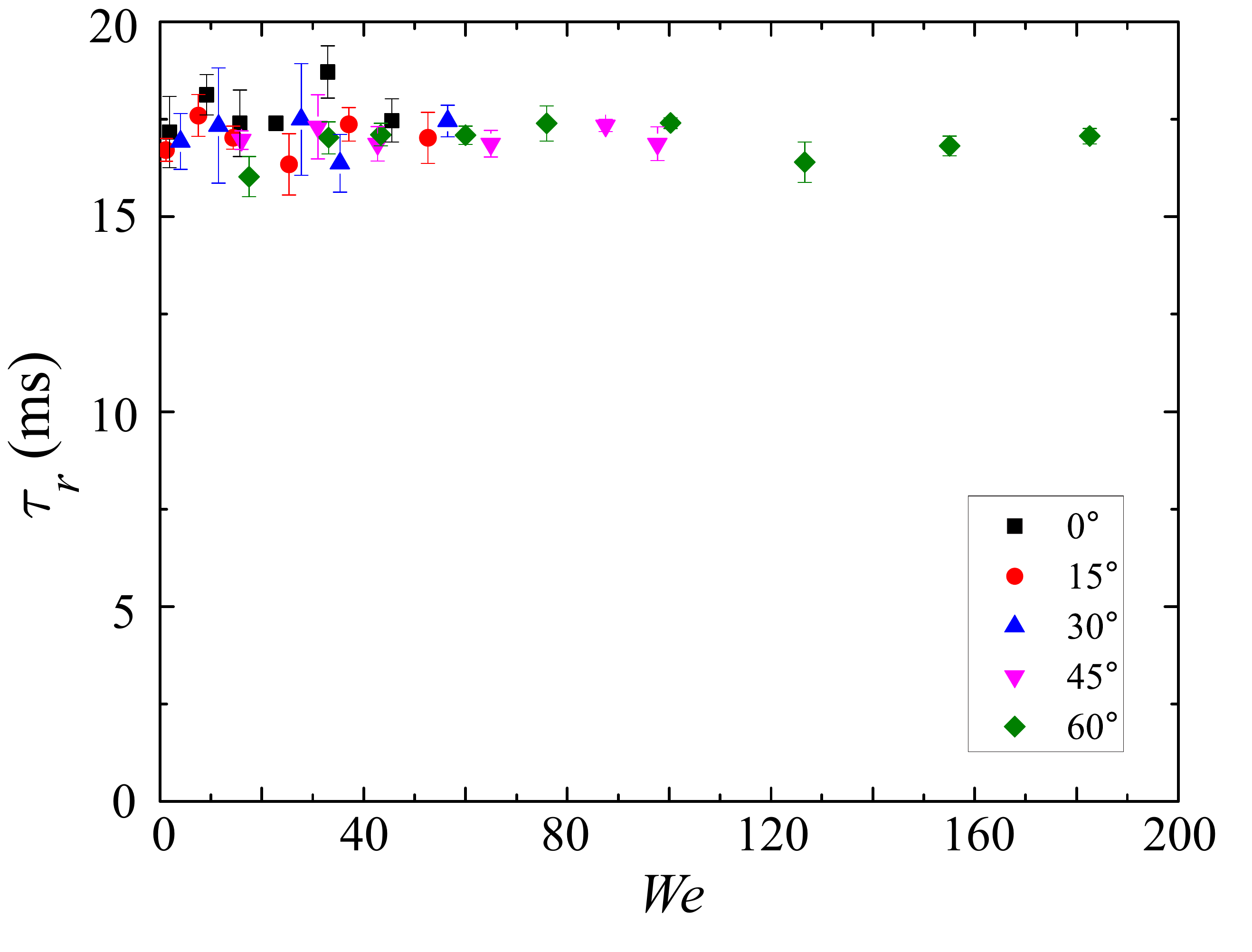}
\caption{Residence time $\tau_r$ of a Leidenfrost drop on substrates with various inclinations versus {\it We}. 
}
\label{time}
\end{figure}

In our experiments, the drop size $D_0$ is a constant about 2.1 mm ($\pm$ 0.1 mm) and the corresponding $\tau_0\simeq15.85$ ms. $\rho_b$ and $\sigma_b$ are the density and the surface tension of ethanol at the boiling point. When the impact velocity is limited to the regime without the splashing formation, the residence time $\tau_r$ versus {\it We} for a Leidenfrost drop is shown in Fig.~\ref{time}. For different impact velocities and inclinations, the residence time of a Leidenfrost drop is nearly a constant of 17 ms, and $\tau_r/\tau_0\simeq1.07$, which is also observed in T. Tran {\it et al}~\cite{Tran2013softmatter}. It suggests that the capillary oscillation determines the residence time of a Leidenfrost drop impact on inclined substrates.

\subsection{Leidenfrost regime: sliding distance} \label{exp3}

In previous sections, we draw the conclusion that the spreading dynamics is strongly determined by the normal velocity of the Leidenfrost drop. In this section, we focus on the effect of the parallel velocity of an impacting Leidenfrost drop. As shown in Fig.~\ref{entire}, besides the spreading and retracting processes, the drop slides along the substrate due to the parallel velocity. In addition, the existence of the vapor layer prevents the drop from directly contacting the substrate. Then the shear stresses acting on the drop bottom can be neglected. 

\begin{figure}
\centering
\includegraphics[width=0.9 \linewidth]{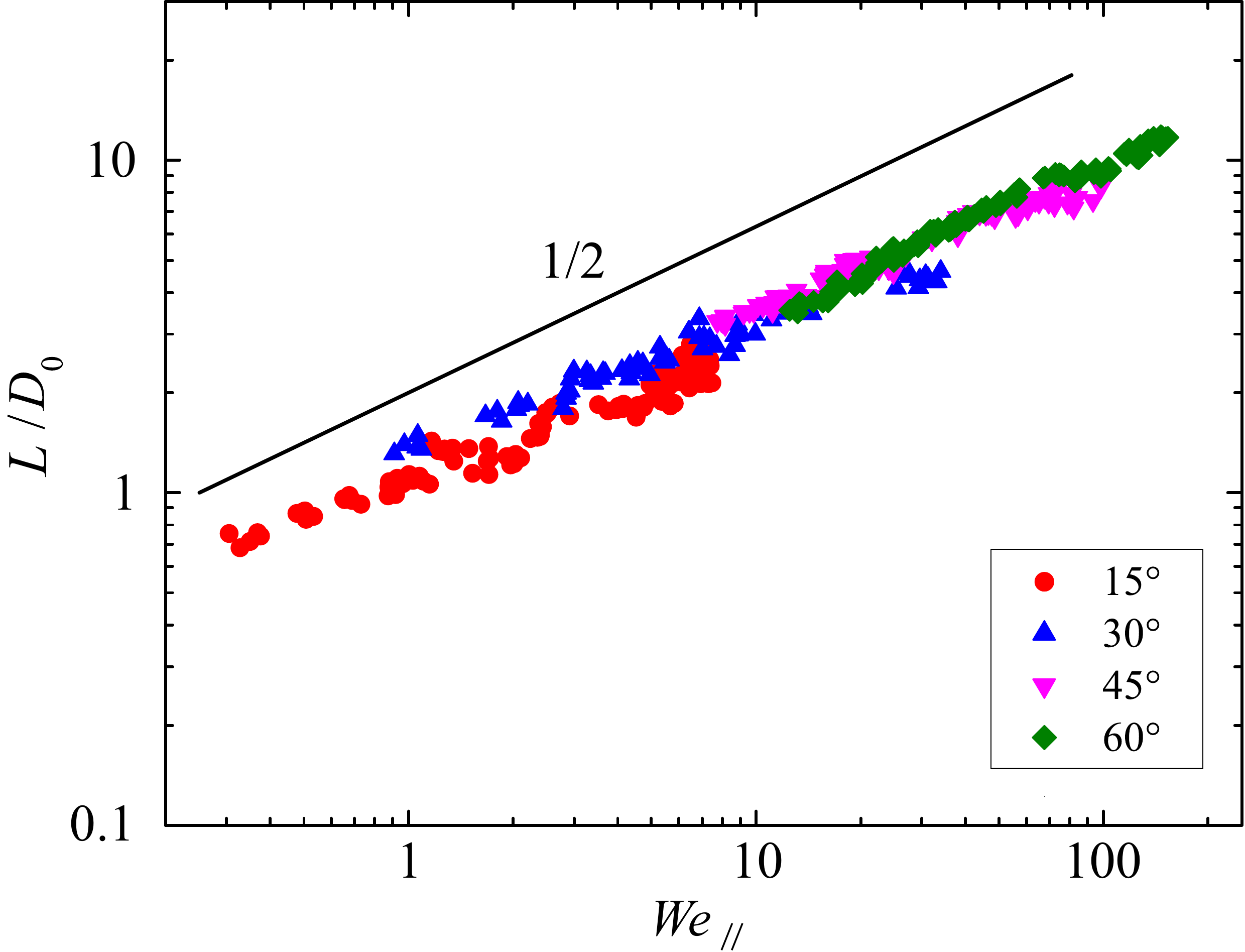}
\caption{Dimensionless sliding distance $L/D_0$ versus parallel Weber number {\it We}$_\sslash$ for different inclinations. The solid line represents the scaling law in Eqn.~\ref{L}, $L/D_0 \propto We_\sslash^{1/2}$, resulting from the free-slip boundary condition.}
\label{slide}
\end{figure}

To quantitatively analyze the sliding behaviors, we define the sliding distance $L$ as the length that a drop moves along the substrate during the residence time. In Fig.~\ref{slide}, we show a log-log plot of the dimensionless slide distance $L/D_0$ as a function of the parallel Weber number for different inclined substrates. The parallel Weber number {\it We}$_\sslash$ is defined as
\begin{equation}
We_\sslash=\dfrac{\rho D_0 U_\sslash^2}{\sigma},
\end{equation}
where the parallel velocity $U_\sslash = U\sin\alpha$. 

All data collapse to a master curve for different inclinations. The scaling of the dimensionless slide distance $L/D_0$ is close to the 1/2-scaling of the {\it We}$_\sslash$, which holds for the free-slip surface assumption and neglects of the gravitational acceleration. On the basis of the free-slip boundary condition, the sliding distance can be expressed as $L \propto \tau_r U_\sslash$. Because $\tau_r$  is nearly a constant, $L/D_0$ can be expressed in terms of the parallel Weber number as:

\begin{equation}
L/\emph{D}_0 \propto We_\sslash^{1/2}.   \label{L}
\end{equation}

Thus, we conclude that the parallel velocity component contributes to the sliding process of a Leidenfrost drop on the substrate. The drop dynamics can be analyzed by decomposing the impact velocity into normal velocity and parallel velocity, which independently influence the spreading process and sliding process respectively.


\subsection{Splashing criterion}  \label{exp4}

\begin{figure}
\centering
\includegraphics[width=0.9 \linewidth]{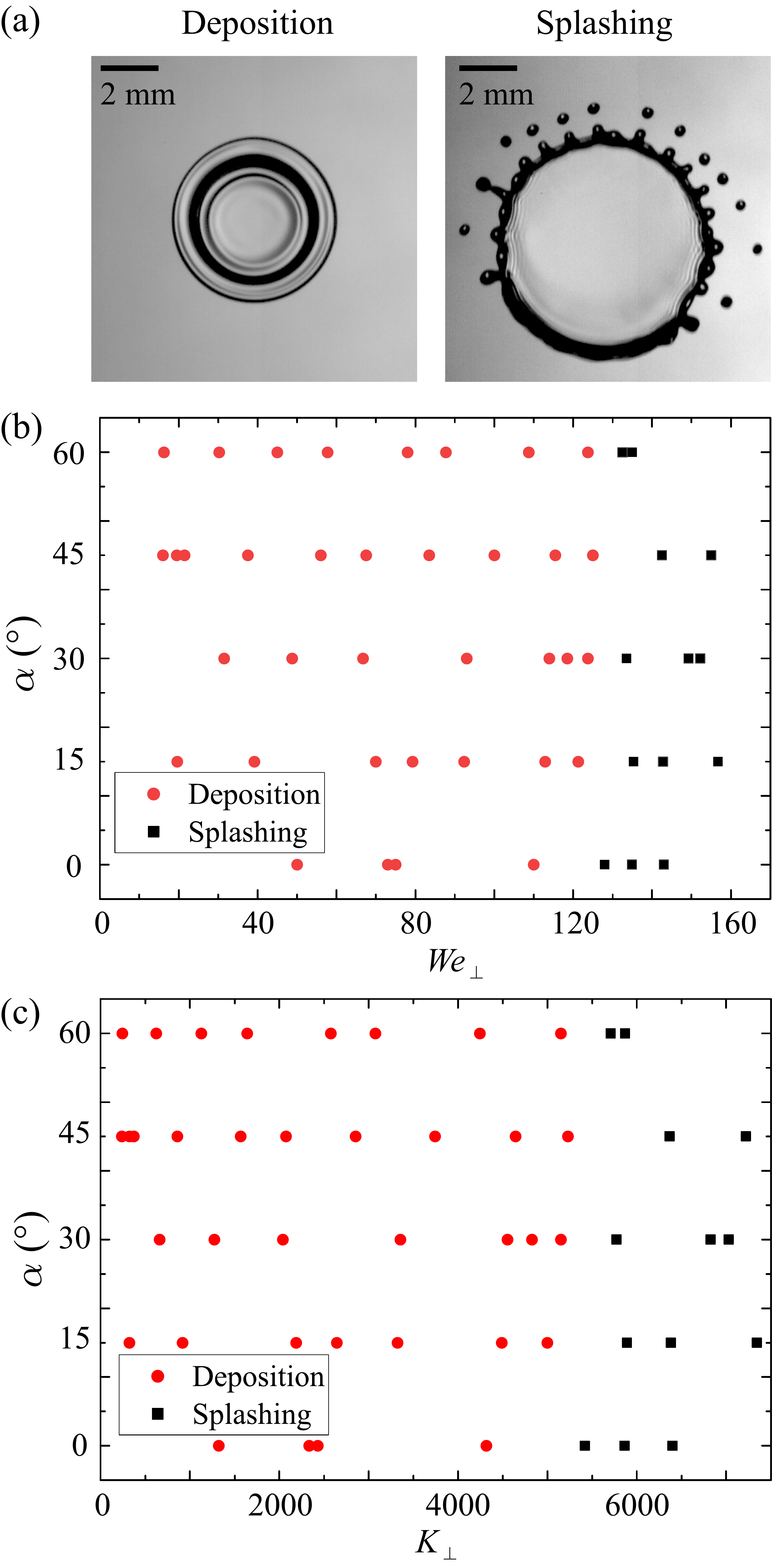}
\caption{The critical splashing criterion. (a) Top-view observations of the deposition and the splashing states. Scale bars represent 2 mm. (b) A phase diagram of the drop state versus {\it We}$_\perp $ for various inclinations is plotted to find the critical splashing criterion. The critical splashing criterion {\it We}$_\perp ^*\simeq$ 120. (c) The critical splashing criterion is $K_\perp  = We_\perp Re_\perp^{1/2} \simeq$ 5300.}
\label{splash}
\end{figure}

When the impact velocity of a drop further increases, the surface tension can not keep the drop integral any more. As shown in the right image of Fig.~\ref{splash}(a), the drop may fragment into tiny drops violently ejecting radially outwards, which is called the splashing phenomenon. When the drop can still keep integral and no tiny drops occur, we call it the deposition state (see the left image of Fig.~\ref{splash}(a)). 

A drop impacts on inclined substrates at room temperature is asymmetrical, and the parallel velocity component also affects the splashing threshold~\cite{Bird2009}. However, in our experiments, due to the high normal impact velocity in the splashing state, the Leidenfrost drop always keeps symmetrical before splashing. And the drop splashing criterion is mainly controlled by the normal velocity component\cite{hao2019droplet}, a phase diagram of the drop state versus {\it We}$_\perp $ is plotted to find the critical splashing criterion. In Fig.~\ref{splash}(b), the drop stays in the deposition state for the small {\it We}$_\perp $. When {\it We}$_\perp $ increases above a critical value, the drop enters the splashing state. And the critical {\it We}$_\perp ^* \simeq120$ is nearly independent on the substrate inclinations, which agrees well with the model proposed by G. Riboux et al~\cite{Riboux2016Maximum}. To date, it is found that the splashing threshold is also dependent on experimental conditions such as the surrounding pressure~\cite{tsai2010micropatterns, xu2005drop}, substrate roughness~\cite{tsai2010micropatterns, kim2014drop}, and substrate temperature~\cite{Riboux2016Maximum}. 

 Another quantity on the splashing threshold is the splashing parameter $K_\perp = We_\perp Re_\perp^{1/2}$ ~\cite{thoroddsen2011droplet}. In Fig.~\ref{splash}(c), we find this normal splashing parameter $K_\perp  \simeq$ 5300 is nearly a constant for different inclinations in the current parameter regime. The existence of the vapor layer prevents the Leidenfrost drop from directly contacting the substrate, which results in the different splashing threshold from that of drop impact on substrates at room temperature~\cite{Bird2009}.

\section{Discussion and Conclusions} \label{con}
In this paper, we investigate the impact of ethanol drops on horizontal and inclined superheated substrates. For horizontal and inclined substrates, there are three different boiling regimes. We use the bottom view to get the accurate dynamic Leidenfrost temperature for various inclinations. The dynamic Leidenfrost temperature increases with {\it We}. For a fixed {\it We}, the dynamic Leidenfrost temperature decreases with the inclination. And the dynamic Leidenfrost temperature is mainly determined by the normal Weber number {\it We}$_\perp $. 

We also investigate the spreading and retracting dynamics of Leidenfrost drops. During the spreading process, drops spread uniformly even for different inclinations. However, in the retracting process, drops with low initial velocities tend to retract slowly in the sliding direction. But as the impact velocity increases, the drop retracts more and more uniformly. The dimensionless maximum spreading factor $D_m/D_0$ is mainly determined by the normal Weber number {\it We}$_\perp $. And it fits well with the model proposed by Wildeman et al~\cite{wildeman2016spreading} at high {\it We}$_\perp $, which assumes the boundary condition of the drop is free-sliding. The residence time of the drop is approximately related to the capillary time of a drop, which is nearly a constant and independent of impact velocity and inclination. Besides the spreading and retracting process, the drop also slides along the substrate. The dimensionless slide distance $L/D_0$ is influenced by the parallel component of the impact velocity, which scales with the parallel Weber number {\it We}$_\sslash^{1/2}$. 

When we further increase the impact velocity, the drop fragments into many tiny droplets, which is called the splashing phenomenon. The critical splashing criterion is found to be {\it We}$_\perp ^*\simeq$ 120 or $K_\perp  \simeq$ 5300.

Our study decomposes movements of the Leidenfrost drop into two directions, the normal and the parallel directions to the substrate. We find that the normal {\it We}$_\perp $ mainly determines the spreading and retracting processes of the drop, and the parallel {\it We}$_\sslash$ determines the sliding process. Because the vapor layer can prevent the drop from directly contacting the substrate, these two processes in orthogonal directions are nearly independent. Then we can change the spreading diameter and the sliding distance by easily changing the inclination, which can help us to adapt the range of various parameters in industrial applications. 
In addition, an inclined superheated substrate is a very convenient tool to change the shape and the moving direction of a drop. After bouncing from the substrate, the sphere drop becomes nearly cylindrical, which has a larger surface area. It can be used in many applications, such as improving the combustion efficiency of fuel sprays, improving the reaction efficiency of chemical reagents, changing the mark of the spray coating, and so on.

\begin{acknowledgments}
We gratefully acknowledge D. Lohse, H. A. Stone and H. Xu for fruitful discussions. This work is supported by the Natural Science Foundation of China under grant nos. 11861131005 and 11988102.
\end{acknowledgments}

%


\end{document}